\documentclass[preprint,aps,nofootinbib,superscriptaddress,showpacs,showkeys]{revtex4}
\usepackage{epsfig}
\usepackage{amssymb}
\usepackage{amsmath}
\usepackage{bm}
\usepackage{hyperref}

\begin{document}

\title{GEANT4 and PHITS simulations of the shielding of neutrons from $^{252}$Cf source}

\author{Jae Won Shin}
\affiliation{Department of Physics, Sungkyunkwan University,
Suwon 440-746, Korea}

\author{Sang-In Bak}
\affiliation{Department of Energy Science, Sungkyunkwan University,
Suwon 440-746, Korea}

\author{Doyoon Kim}
\affiliation{Department of Energy Science, Sungkyunkwan University,
Suwon 440-746, Korea}

\author{Chong Yeal Kim}
\affiliation{Department of Radiation Science $\&$ Technology, Chonbuk National University, Jeonju 561-756, Korea}

\author{Seung-Woo Hong}
\email{swhong@skku.ac.kr}
\affiliation{Department of Physics, Sungkyunkwan University,
Suwon 440-746, Korea}

\date{26 March 2014}

\begin{abstract}

Monte Carlo simulations by using GEANT4 and PHITS are performed for 
studying neutron shielding abilities of several materials, 
such as graphite, iron, polyethylene, NS-4-FR and KRAFTON-HB.
As a neutron source $^{252}$Cf is considered.
For the Monte Carlo simulations by using GEANT4, 
high precision (G4HP) models with G4NDL 4.2 based on ENDF/B-VII data are used.
For the simulations by using PHITS, JENDL-4.0 library are used.
The neutron dose equivalent rates with 
or without five different shielding materials are 
estimated and compared with the experimental values.
It is found that the differences between the shielding abilities calculated 
by using GEANT4 with G4NDL 4.2 and 
PHITS with JENDL-4.0 library are not significant 
for all the cases considered in this work.
We investigate the accuracy of the neutron dose equivalent rates 
obtained by GEANT4 and PHITS 
by comparing our simulation results with experimental data 
and other values calculated earlier.
The calculated neutron dose equivalent rates agree well 
with the experimental dose equivalent rates within 
20\% errors except for polyethylene.
For polyethylene material, discrepancy between 
our calculations and the experiments are up to 40\%, 
but all simulations show consistent features.

\end{abstract}



\maketitle

\section{INTRODUCTION}

Accurate estimations of the neutron shielding abilities are essential 
for safety requirement in the design of facilities 
such as accelerators and nuclear reactors 
or a shielding container of neutron emitting sources.
In determining the shielding abilities, 
characteristics of the materials is a major factor.
Many studies have been done for 
neutron shielding abilities for various 
shielding materials \cite{ueki1, ueki2, ueki3, ueki4}, 
shielding designs \cite{int0, int1, phits_ex_shield3}, 
further development of new neutron shielding materials \cite{int3, int4} and etc.
Also, benchmark simulations by using different Monte Carlo codes 
\cite{bench1, bench2, bench3, bench4, bench5, bench6, bench7, bench8} 
have been done with MCNP, MCNP4B2, SAS, SCALE and GEANT4.

Effectiveness of various shielding materials 
for neutrons from a $^{252}$Cf source 
were experimentally evaluated in \cite{ueki4}.
In Ref. \cite{bench8}, 
the neutron shielding was studied with GEANT4 code \cite{g4n1}.
For neutron interactions, both high precision (G4HP) models 
with G4Neutron Data Library (G4NDL) 3.13 
based on ENDF/B-VI library 
and low energy parameterized (G4LEP) models were used and tested. 
Relative neutron dose equivalent rates were calculated and 
the results were compared with the experimental data \cite{ueki4}.
It was shown that G4HP models were a good candidate for accurate 
simulations of neutron shieldings.

As an extension of these previous studies 
\cite{bench1, bench2, bench3, bench4, bench5, bench6, bench7, bench8}, 
we have performed in this work Monte Carlo simulations for neutron shielding 
by using GEANT4 and PHITS \cite{phits1, phits2, phits3}.
In Ref. \cite{bench8} G4NDL 3.13 based on ENDF/B-VI was used for GEANT4 v9.3, 
but the latest version of nuclear data available for GEANT4 is now G4NDL 4.X 
based on ENDF/B-VII.
In this work, we also used different Monte Carlo code PHITS 
for a benchmark purpose.
For neutron interactions, G4HP models with G4NDL 4.2 based on ENDF/B-VII library 
and JENDL-4.0 library are used 
for GEANT4 v9.6 and PHITS v2.52 simulations, respectively.
As a neutron source, $^{252}$Cf is assumed.
$^{252}$Cf emits neutrons with 
an average energy of $\sim$ 2.2 MeV by spontaneous fission.
In Ref. \cite{ueki3, ueki4} the neutron dose equivalent rates are extracted 
instead of measuring neutron spectrum.
Thus, we also estimated the neutron dose equivalent rates for comparison with 
the experimental results \cite{ueki3, ueki4}.
The neutron dose equivalent rates with and without the shield are estimated for 
shielding materials such as graphite, iron, polyethylene, NS-4-FR and KRAFTON-HB. 
Graphite is used in reactor design as a moderator and reflector.
Iron is commonly used for the high energy neutron shielding in accelerator facilities.
Polyethylene is a popular neutron shielding material that highly contains hydrogen.
NS-4-FR is an epoxy resin containing heavy elements together with boron 
to reduce the production of secondary $\gamma$-rays due to 
thermal neutron absorption.
KRAFTON-HB also contains boron for the same purpose, 
and was developed as an advanced
shielding material for fast breeder reactors.
By comparing our results with the experimental data \cite{ueki3, ueki4}
and the previous simulation results \cite{bench1, bench3, bench5}, 
we can compare the accuracy of the neutron dose equivalent rates obtained by GEANT4 
and PHITS with new nuclear data libraries.

The outline of the paper is as follows. 
In Sec. II, simulation tools and simulation set up are described.
In Sec. III, the calculated neutron energy distributions 
scored in the detector region and 
the corresponding neutron dose equivalent rates are shown.
The resulting neutron dose equivalent rates 
are compared with experimental data.
A summary is given in Sec. IV.

\section{Method}

\subsection{Simulation tools}

As Monte Carlo simulation tools, 
we use GEANT4 and PHITS and compare the results from them.
For the low energy neutron interactions in material, 
G4HP models with G4NDL based on ENDF library \cite{endf} 
and JENDL library \cite{jendl} are used in 
GEANT4 and PHITS simulations, respectively.
Here we list some key features of the simulation tools.

GEANT4 : GEANT4 (GEometry ANd Tracking) is a simulation tool kit written in C++
language, which allows microscopic simulations of the
propagation of particles interacting with materials.
It is being widely used in many different fields, 
such as neutron shielding studies \cite{bench8, G4shield2}, 
medical physics \cite{G4Med1, G4Med2, Shin1, Shin1a}, 
accelerator based single event upset studies \cite{ShinAcc1, Shin2},
environment radiation detection studies \cite{G4Det2, G4Det3, Shin4}, and etc.

In our previous work \cite{bench8}, 
we showed that G4HP models with G4NDL 3.13 were better than 
G4LEP models.
For this reason, 
G4HP models with G4NDL 4.2 are used in this work.
Both G4HP models and G4LEP models include 
cross sections for elastic, inelastic scattering,
capture, fission and isotope production.
The energy range of these classes are from 
thermal energies to 20 MeV.
Data of G4NDL 4.2 come largely from the ENDF/B-VII library.
The validations and the detailed descriptions of the GEANT4, 
G4HP models and G4NDL can be found on the GEANT4 website \cite{g4_web}.

PHITS : PHITS (Particle and Heavy-Ion Transport code System)
is a multi-purpose Monte Carlo 
transport code system for heavy ions and all particles 
with the energies from meV up to 200 GeV.
It was developed by the collaboration of 
Japan Atomic Energy Agency (JAEA), 
Research Organization for Information Science Technology (RIST), 
High Energy Accelerator Research Organization (KEK), 
and several other institutes.
PHITS is also used for many scientific studies such as
space technology \cite{phits_ex_space1, phits_ex_space2},
medical physics \cite{phits_ex_mp1, phits_ex_mp2, phits_ex_mp3},
shielding designs \cite{phits_ex_shield1, phits_ex_shield2, phits_ex_shield3},
accelerator applications \cite{phits_ex_acc1, phits_ex_acc2} and etc.

Neutron simulations can be done by using 
various evaluated nuclear data libraries in PHITS code.
JENDL-4.0 library, which is the latest version of JENDL, 
is used in this work.
Additional information and detailed descriptions
of PHITS can be found on the web \cite{phits_web}.

\subsection{Simulation set up}

In this work, we consider the neutron shielding experiments 
in Ref. \cite{ueki3, ueki4}, 
where the experiments were done to evaluate the
effectiveness of various shielding materials for neutrons from a $^{252}$Cf source.
$^{252}$Cf emits neutrons with an average energy of $\sim$2.2 MeV.
For the generation of neutrons from $^{252}$Cf source, 
Watt fission spectrum \cite{watt_1, watt_2, x5}
are used in our simulations.
The spontaneous fission neutron spectrum 
given by the Watt fission spectrum is expressed as

\begin{eqnarray}
f(E) = {\rm exp}(-\frac{E}{1.025}){\rm sinh}(2.926E)^{1/2},
\end{eqnarray}
where {\em E} is the neutron energy in MeV \cite{x5}.

\begin{figure}[tbp]
\epsfig{file=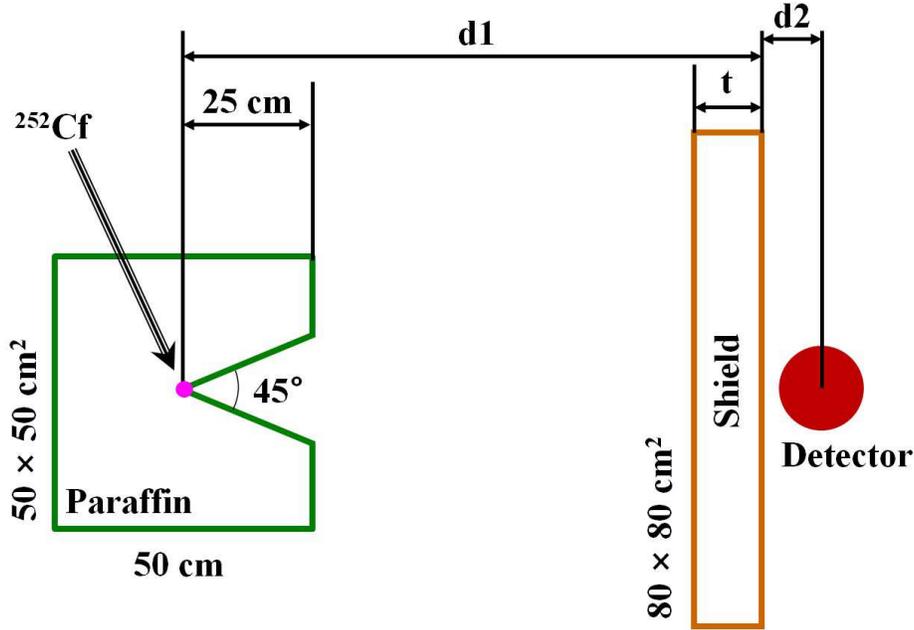, width=5in}
\caption{Schematic diagram showing the simulation geometry.}
\label{figure1}
\end{figure}
The geometry of the shield and the detector is 
drawn in Fig. \ref{figure1}. 
The $^{252}$Cf source is surrounded by a 
50 $\times$ 50 $\times$ 50 cm$^{3}$ paraffin container block.
There is a conical shape of an opening in the paraffin 
so that the neutrons from the $^{252}$Cf can pass freely and propagate 
through the air to reach the detector behind the shield.
The neutron detector has a cylindrical shape of radius 5.25 cm 
and length 10 cm. 
It is assumed that all the neutrons that reach the detector are scored.
Shielding materials, source strengths of $^{252}$Cf, d1, d2 and 
the thicknesses (t) of the shielding materials are 
tabulated in Table \ref{table1}.
Five different shielding materials such as graphite, iron, polyethylene, NS-4-FR 
and KRAFTON-HB are considered.
The components of these materials and their mass fractions are 
tabulated in Table \ref{table2}.

%%%%%%%%%%%%%%%%%%%%%%%%%%%%%%%%%%%%%%%%%%%%%%%%%%%%%%%%%%%%%%%%%%%%%%%%%%%%%
\begin{table}
\caption{Parameters for the neutron shielding experiments and simulations used in this work}
\begin{ruledtabular}
\begin{tabular}{cccccc}
materials & neutron source strength (n/sec) & d1 (cm) & d2 (cm) & t (cm)\\
\colrule
 Graphite \cite{ueki3}& 1.62 $\times$ 10$^{7}$ & 90 & 20 & 5, 15, 25, 35 \\
 Iron \cite{ueki3}& 1.50 $\times$ 10$^{7}$ & 90 & 20 & 5, 15, 25, 35\\
 Polyethylene \cite{ueki3}& 6.28 $\times$ 10$^{7}$ & 90 & 20 & 5, 15, 25, 35\\
 NS-4-FR \cite{ueki4}& 5.45 $\times$ 10$^{7}$ & 100 & 15 & 5, 10, 15, 20, 25\\
 KRAFTON-HB \cite{ueki4}& 5.45 $\times$ 10$^{7}$ & 100 & 15 & 5.3, 10.6, 15.9, 21.2, 26.5, 31.8\\
\end{tabular}
\end{ruledtabular}
\label{table1}
\end{table}
%%%%%%%%%%%%%%%%%%%%%%%%%%%%%%%%%%%%%%%%%%%%%%%%%%%%%%%%%%%%%%%%%%%%%%%%%%%%%

\begin{figure}[tbp]
\epsfig{file=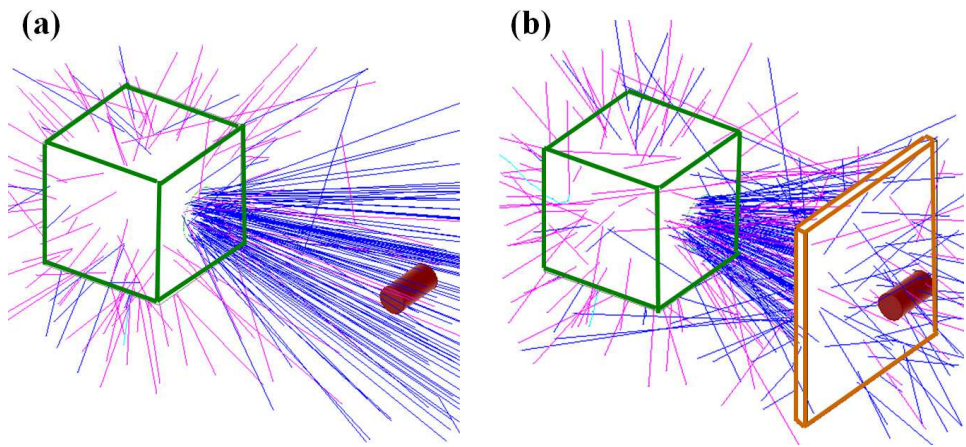, width=5in}
\caption{(Color online) OpenGL pictures showing the simulation geometry 
and the neutrons (blue) and gammas (pink) without (a) and with (b) the shielding.}
\label{figure2}
\end{figure}

Typical simulation snap shots drawn by using OpenGL library 
are shown in Fig. \ref{figure2}.
Figures \ref{figure2} (a) and \ref{figure2} (b) show the propagation of 
neutrons and gammas without and with the shielding material, respectively.
It can be seen that a large number of neutrons fly through the opening 
but are mostly blocked by the shielding material.
With this geometry, the energy distributions of the neutrons 
scored in the detector after escaping through
the shielding materials are calculated.

%%%%%%%%%%%%%%%%%%%%%%%%%%%%%%%%%%%%%%%%%%%%%%%%%%%%%%%%%%%%%%%%%%%%%%%%%%%%%
\begin{table}
\caption{Components and mass fractions of shielding materials used in this work}
\begin{ruledtabular}
\begin{tabular}{cccccc}
       & graphite & iron  & polyethylene & NS-4-FR & KRAFTON-HB \\
\colrule
H      &     -    &   -   &     14.4     &   5.92  & 10.66 \\
B      &     -    &   -   &       -      &   0.94  & 0.78 \\
C      &    100   &   -   &     85.6     &   27.63 & 75.29 \\
N      &     -    &   -   &       -      &   1.98  & 2.20 \\
O      &     -    &   -   &       -      &   42.29 & 10.69 \\
Al     &     -    &   -   &       -      &   21.24 & - \\
Si     &     -    &   -   &       -      &    -    & 0.38 \\
Fe     &     -    &  100  &       -      &    -    & - \\
\end{tabular}
\end{ruledtabular}
\label{table2}
\end{table}
%%%%%%%%%%%%%%%%%%%%%%%%%%%%%%%%%%%%%%%%%%%%%%%%%%%%%%%%%%%%%%%%%%%%%%%%%%%%%

%%%%%%%%%%%%%%%%%%%%%%%%%%%%%%%%%%%%%%%%%%%%%%%%%%%%%%%%%%%%%%%%%%%%%%%%%%%%%
%%%%%%%%%%%%%%%%%%%%%%%%%%%%%%%%%%%%%%%%%%%%%%%%%%%%%%%%%%%%%%%%%%%%%%%%%%%%%

To make a comparison with the experimental results given in terms of 
dose equivalent rates \cite{ueki3, ueki4},
we need to calculate dose equivalent rates.
To estimate the human biological dose equivalent rates, 
one often uses a conversion factor which converts the neutron 
flux to human biological dose equivalent rate. 
In this work, neutron dose equivalent rates are calculated by using the 
flux to dose conversion factor from the
National Council on Radiation Protection and 
Measurements (NCRP-38) standard \cite{ncrp}.

\section{Results}

Simulations are performed by using GEANT4 v9.6 with G4NDL 4.2 
and PHITS v2.52 with JENDL-4.0.
First, the calculated neutron energy spectra scored in the detector region 
for various shielding materials are shown.
The scored numbers of neutrons are normalized by 
the number of neutrons generated by the source.
We refer to these values as "counts".
In Ref. \cite{bench8}, the neutron shielding simulations were done by using 
GEANT4 v9.3 with G4NDL 3.13.
The data of G4NDL 3.13 are based on the ENDF/B-VI library.
For the comparison between the results of Ref. \cite{bench8} and the present work, 
the calculations by using GEANT4 v9.3 with G4NDL 3.13 
have been also performed and the results are compared.
The results from GEANT4 v9.6 with G4NDL 4.2 
and PHITS v2.52 with JENDL-4.0 are also compared with each other.

The calculated the neutron dose equivalent rates by using 
the flux to dose conversion factor from NCRP-38 are compared with 
those from the experiments \cite{ueki3, ueki4} 
and other values calculated earlier \cite{bench1, bench3, bench5}.

\subsection{Neutron spectra scored in the detector region}

Neutron spectra scored in the detector region obtained from 
GEANT4 with G4NDL 3.13 and G4NDL 4.2 and 
those from PHITS with JENDL-4.0 are compared with each other.
However, differences among the simulation results 
are not significant for all the cases.
For this reason, we show only the results 
from GEANT4 v9.6 with G4NDL 4.2 and PHITS v2.52 with JENDL-4.0.
For brevity, we refer to the calculations by using 
GEANT4 v9.6 with G4NDL 4.2 and PHITS v2.52 with JENDL-4.0 
as simply GEANT4 and PHITS, respectively.

%%%%%%%%%%%%%%%%%%%%%%%%%%%%%%%%%%%%%%%%%%%%%%%%%%%%%%%%%%%%%%%%%%%%%%%%%%%%%
\begin{figure}[tbp]
\epsfig{file=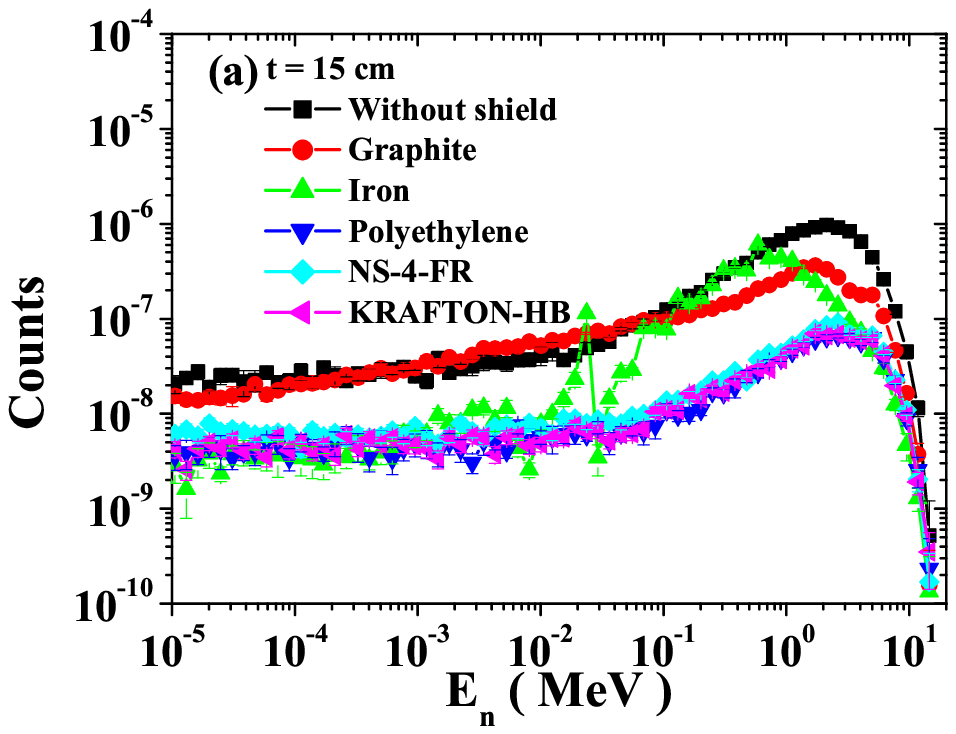, width=4in}
\epsfig{file=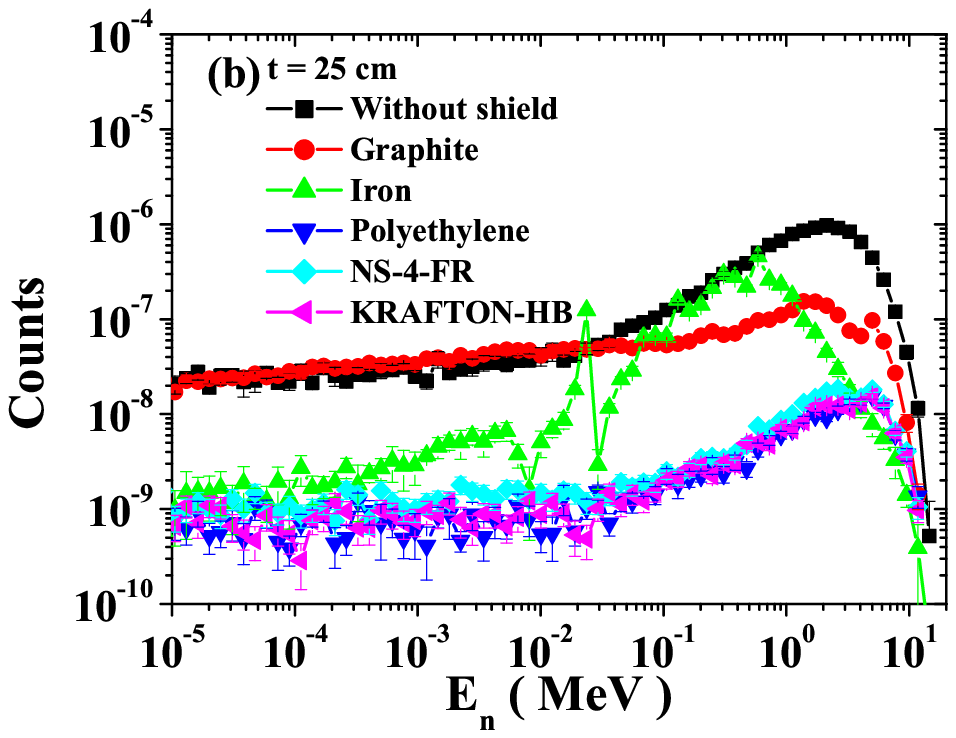, width=4in}
\caption{(Color online) Energy distributions of 
the neutrons scored in the detector region 
for various shielding materials.
(a) and (b) show the results for the 15 and 25 cm of 
the shielding thickness, respectively.}
\label{figure3}
\end{figure}
%%%%%%%%%%%%%%%%%%%%%%%%%%%%%%%%%%%%%%%%%%%%%%%%%%%%%%%%%%%%%%%%%%%%%%%%%%%%%

Figure \ref{figure3} shows the energy distributions of the neutrons 
scored in the detector region after passing through the shielding materials. 
The results for five different shielding materials with 15 and 25 cm of 
thicknesses are plotted in Fig. \ref{figure3} (a) and \ref{figure3} (b), respectively.
It can be seen that graphite and iron reduce the number of neutrons at high energies 
but not at low energies in comparison with the counts without any shielding material. 
Also, graphite and iron do not reduce the neutron flux as much as 
other shielding materials.
For polyethylene, NS-4-FR and KRAFTON-HB, which contain the hydrogen element, 
similar shapes of the spectrum and neutron shielding abilities are observed.
It can be seen that those materials reduce the counts by up to two orders of magnitudes 
in comparison with the counts without any shielding material.

\begin{figure}[tbp]
\epsfig{file=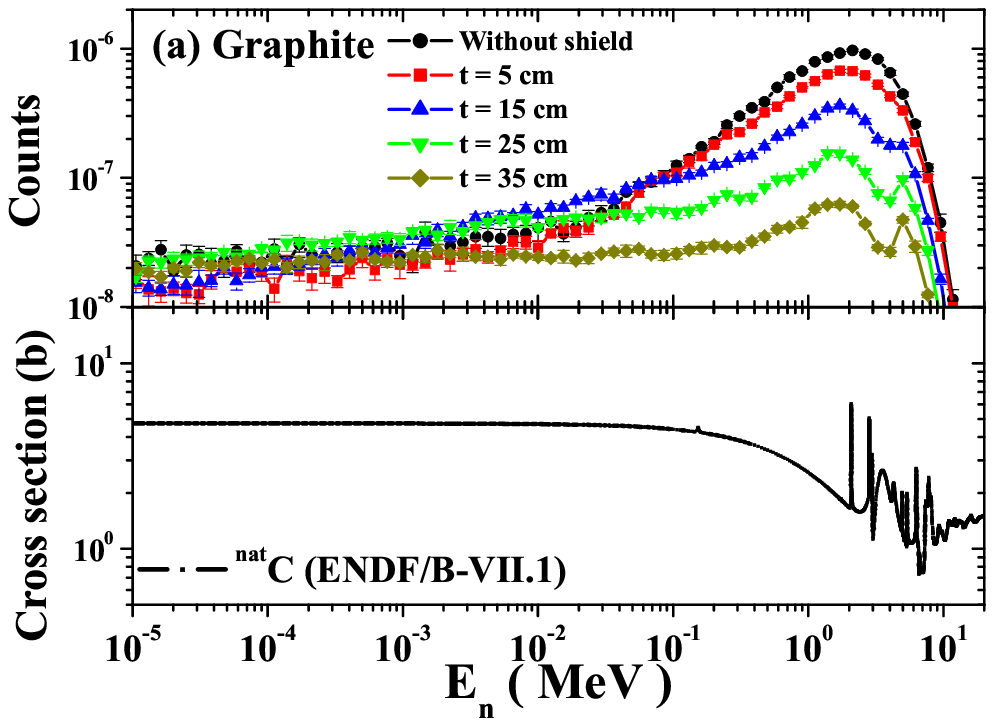, width=4in}
\epsfig{file=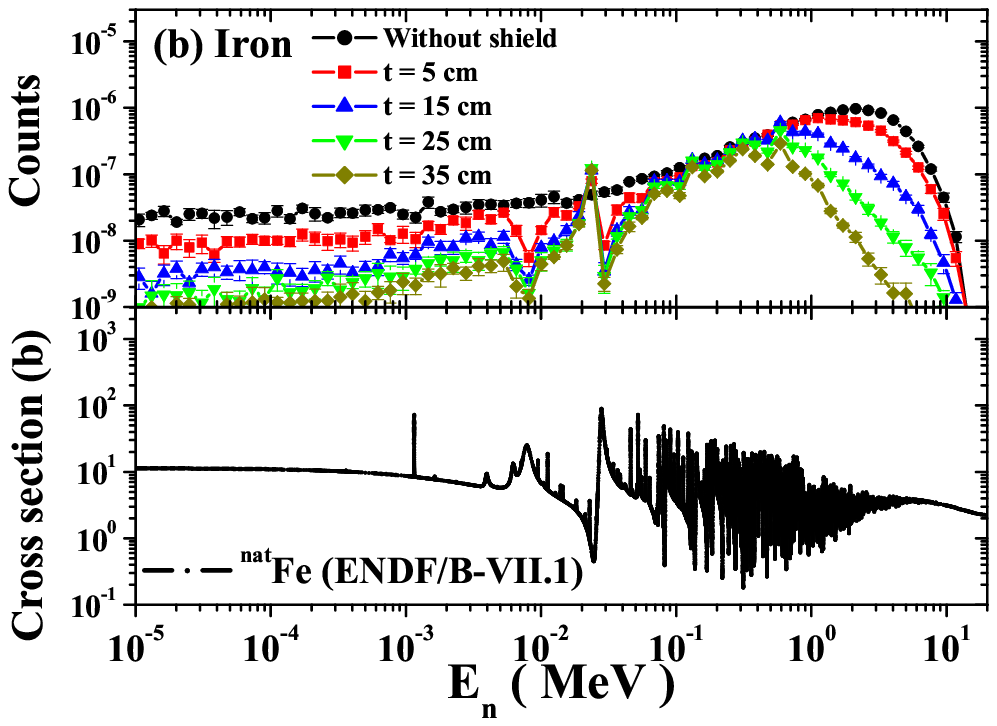, width=4in}
\caption{(Color online) (a) and (b) represent the results 
for graphite and iron shielding materials, respectively.
The upper panels show energy distributions of 
the neutrons in the detector regions and 
the lower panels show the total cross sections of 
the neutron on graphite and iron.
Total cross sections are taken from ENDF/B-VII.1 library \cite{endf}.}
\label{figure4}
\end{figure}

Peak and valley structures are observed for both graphite and iron materials 
in Fig. \ref{figure3}, which can be 
understood from the total cross sections of the neutron on the elements.
Figure \ref{figure4} shows the energy distributions of the neutrons 
and the total cross sections of the neutron on the elements of the shielding materials.
Total cross sections of the neutron on $^{nat}$C, 
$^{54}$Fe, $^{56}$Fe, $^{57}$Fe and $^{58}$Fe 
are taken from ENDF/B-VII.1 \cite{endf}.
Figure \ref{figure4} (a) shows the energy distributions of the neutrons 
scored in the detector region 
with and without graphite materials and the total cross section 
of the neutron on $^{nat}$C.
As the thickness of the graphite material increases, 
the valley of the neutron spectrum near $\sim$3.5 MeV becomes pronounced.
This feature can be understood by the 
total cross section of the neutron on $^{nat}$C plotted in the lower panel. 
A broad peak, which mainly comes from 
the elastic cross section of the neutron on $^{nat}$C, 
is shown near $\sim$3.5 MeV \cite{endf}.
As a result, the neutrons with these energies are 
more reduced than those with other neutrons.

%%%%%%%%%%%%%%%%%%%%%%%%%%%%%%%%%%%%%%%%%%%%%%%%%%%%%%%%%%%%%%%%%%%%%%%%%%%%%
%%%%%%%%%%%%%%%%%%%%%%%%%%%%%%%%%%%%%%%%%%%%%%%%%%%%%%%%%%%%%%%%%%%%%%%%%%%%%

For iron, 
drastic changes are observed in Fig. \ref{figure4} (b).
Two distinct valleys (at $\sim$0.008 and $\sim$0.03 MeV) and one peak ($\sim$0.024 MeV) 
are shown.
One can see that the energy distribution looks similar to a mirror image of 
the total cross section of the neutron on $^{nat}$Fe.
The peaks at $\sim$0.008 and $\sim$0.03 MeV in the lower panel mainly come from 
the contributions from elastic cross section of 
the neutron on $^{54}$Fe and $^{56}$Fe, respectively \cite{endf}.
Also, a very sharp peak at $\sim$0.00115 MeV which comes from 
the contribution of capture cross section of 
the neutron on $^{56}$Fe \cite{endf} is shown in the lower panel.
However, the probability that neutrons reach the energies of those sharp peak 
is very low because the peak has a very narrow width $<$ $\sim$10 eV.
For this reason, 
this sharp peak at $\sim$0.00115 MeV 
does not produce a visible effect on the neutron counts.

%%%%%%%%%%%%%%%%%%%%%%%%%%%%%%%%%%%%%%%%%%%%%%%%%%%%%%%%%%%%%%%%%%%%%%%%%%%%%
\begin{figure}[tbp]
\epsfig{file=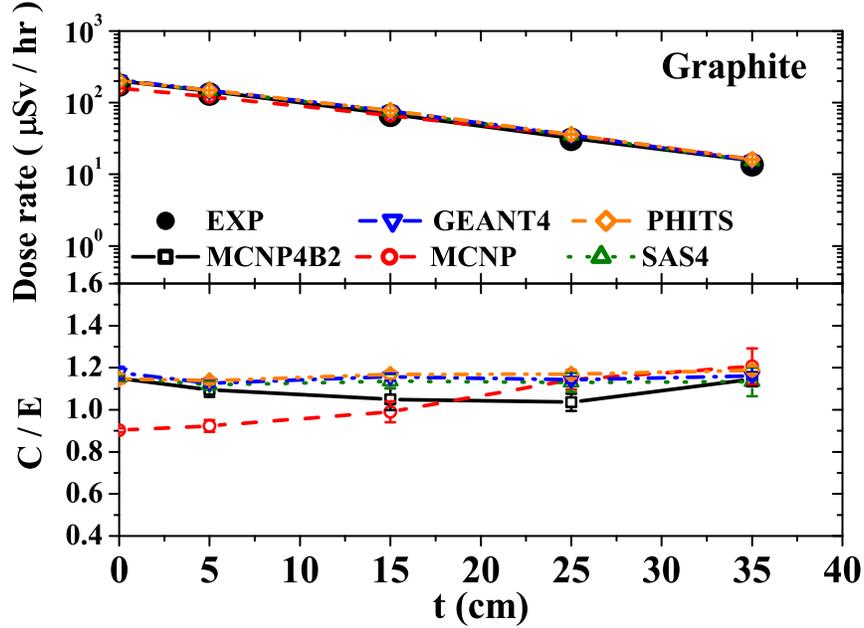, width=5in}
\caption{(Color online) Neutron dose equivalent rates for graphite. 
The upper panel shows the dose equivalent rates in units of $\mu$Sv/hr, and 
the lower panel shows the ratio of the calculated dose equivalent rates 
to the experimental dose equivalent rates.
The filled circles represent the experimental dose equivalent rates \cite{ueki3}.
The open inverted triangles and the diamonds represent 
the calculated dose equivalent rates by using 
GEANT4 and PHITS, respectively.
The open squares, circles and triangles denote 
the calculated dose equivalent rates by using 
MCNP4B2 \cite{bench3}, MCNP \cite{bench1} and SAS4 \cite{bench1}, respectively.
}
\label{figure5}
\end{figure}
%%%%%%%%%%%%%%%%%%%%%%%%%%%%%%%%%%%%%%%%%%%%%%%%%%%%%%%%%%%%%%%%%%%%%%%%%%%%%

%%%%%%%%%%%%%%%%%%%%%%%%%%%%%%%%%%%%%%%%%%%%%%%%%%%%%%%%%%%%%%%%%%%%%%%%%%%%%
\begin{figure}[tbp]
\epsfig{file=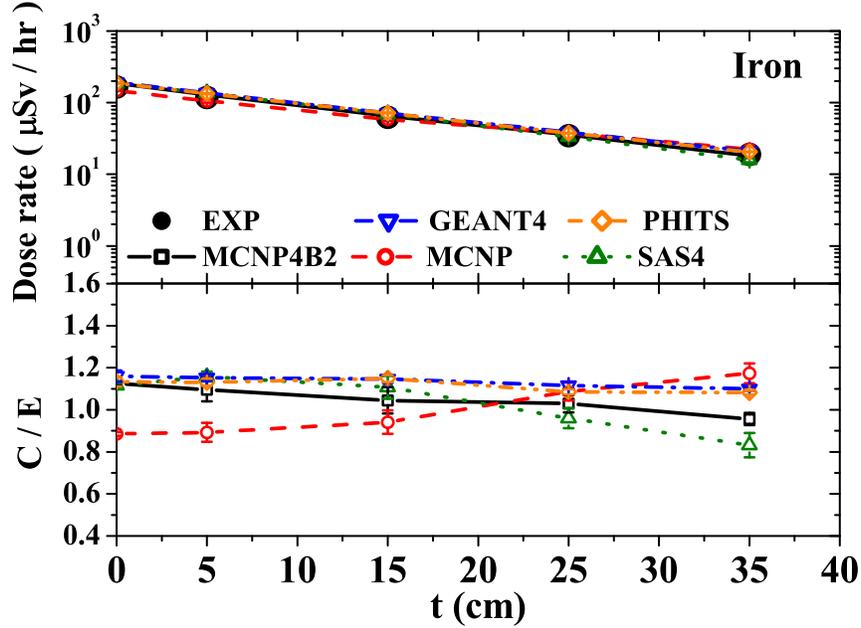, width=5in}
\caption{(Color online) The same as in Fig. \ref{figure5} 
except that the material is iron.}
\label{figure6}
\end{figure}
%%%%%%%%%%%%%%%%%%%%%%%%%%%%%%%%%%%%%%%%%%%%%%%%%%%%%%%%%%%%%%%%%%%%%%%%%%%%%

\subsection{Neutron dose equivalent rates}

With the energy distributions of the neutrons scored in the detector region 
and the conversion factors from NCRP-38, 
we can obtain neutron dose equivalent rates.
Figures \ref{figure5} $\sim$ \ref{figure9} 
show our results for the experimental dose equivalent rates 
for five different shielding materials.
The upper and lower panels of each figure represent 
the dose equivalent rates in units of $\mu$Sv/hr and 
the ratio of the calculated values to 
the experimental values (C/E) \cite{ueki3, ueki4}, respectively.

Figures \ref{figure5} and \ref{figure6} show the neutron dose equivalent rates 
for graphite and iron shielding materials, respectively.
One can see that all the calculations agree with the experiments within $\sim$20\% errors.
C/E from both GEANT4 and PHITS are almost independent of thicknesses 
of the shielding materials.
It means that the calculated neutron dose attenuations are very 
close to those from the experiments.

%%%%%%%%%%%%%%%%%%%%%%%%%%%%%%%%%%%%%%%%%%%%%%%%%%%%%%%%%%%%%%%%%%%%%%%%%%%%%
\begin{figure}[tbp]
\epsfig{file=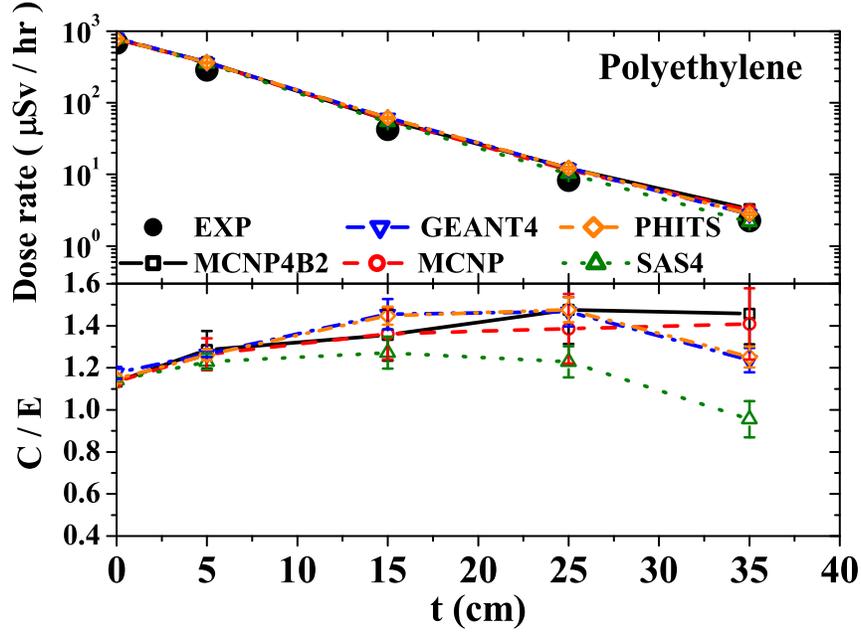, width=5in}
\caption{(Color online) The same as in Fig. \ref{figure5} but for polyethylene.}
\label{figure7}
\end{figure}
%%%%%%%%%%%%%%%%%%%%%%%%%%%%%%%%%%%%%%%%%%%%%%%%%%%%%%%%%%%%%%%%%%%%%%%%%%%%%

For polyethylene, discrepancies between the experimental values 
and the simulations values are up to $\sim$40\% 
as shown in Fig. \ref{figure7}.
All simulations overestimate the experimental dose equivalent rates.
Similar features can be seen in Ref. \cite{bench6}.
In Ref. \cite{bench6}, even though differences 
between the calculated dose equivalent rates and 
the measured one for NS-4-FR and KRAFTON-HB were less than $\sim$20\%,
the calculated dose equivalent rates overestimated the experimental values 
up to $\sim$50\% for polyethylene.

%%%%%%%%%%%%%%%%%%%%%%%%%%%%%%%%%%%%%%%%%%%%%%%%%%%%%%%%%%%%%%%%%%%%%%%%%%%%%
\begin{figure}[tbp]
\epsfig{file=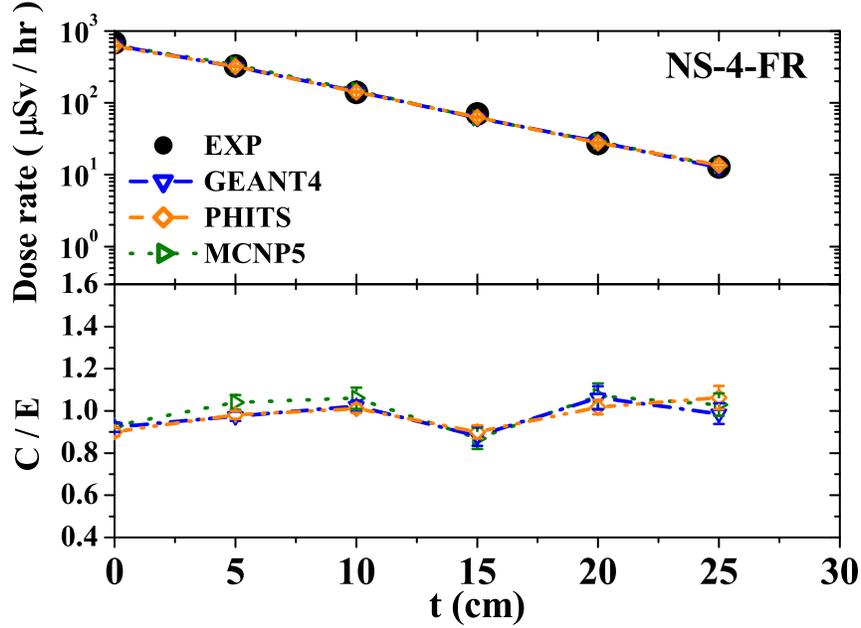, width=5in}
\caption{(Color online) Neutron dose equivalent rates for NS-4-FR. 
The upper panel shows the dose equivalent rates in units of $\mu$Sv/hr, and 
the lower panel shows the ratio of the calculated dose equivalent rates 
to the experimental dose equivalent rates.
The filled circles represent the experimental dose equivalent rates \cite{ueki4}.
The open inverted triangles and diamonds represent 
the calculated dose equivalent rates by using 
GEANT4 and PHITS, respectively.
The open triangles denote the calculated dose equivalent rates by using 
MCNP5 \cite{bench5}.
}
\label{figure8}
\end{figure}
%%%%%%%%%%%%%%%%%%%%%%%%%%%%%%%%%%%%%%%%%%%%%%%%%%%%%%%%%%%%%%%%%%%%%%%%%%%%%

%%%%%%%%%%%%%%%%%%%%%%%%%%%%%%%%%%%%%%%%%%%%%%%%%%%%%%%%%%%%%%%%%%%%%%%%%%%%%
\begin{figure}[tbp]
\epsfig{file=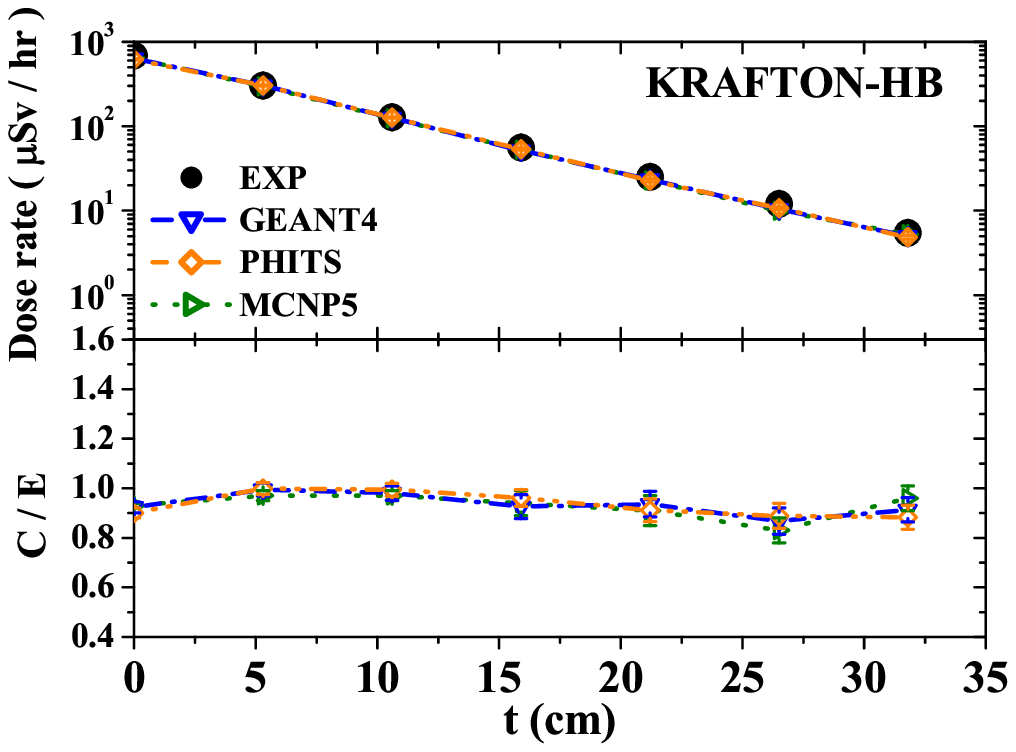, width=5in}
\caption{(Color online) The same as in Fig. \ref{figure8} but for KRAFTON-HB.}
\label{figure9}
\end{figure}
%%%%%%%%%%%%%%%%%%%%%%%%%%%%%%%%%%%%%%%%%%%%%%%%%%%%%%%%%%%%%%%%%%%%%%%%%%%%%

Neutron dose equivalent rates for NS-4-FR and KRAFTON-HB are plotted in 
Fig. \ref{figure8} and Fig. \ref{figure9}, respectively.
One can see that the calculated neutron dose equivalent rates agree well 
with the experimental dose equivalent rates within $\sim$10\% errors.
Also, differences among the calculations are not significant.

\begin{figure}[tbp]
\epsfig{file=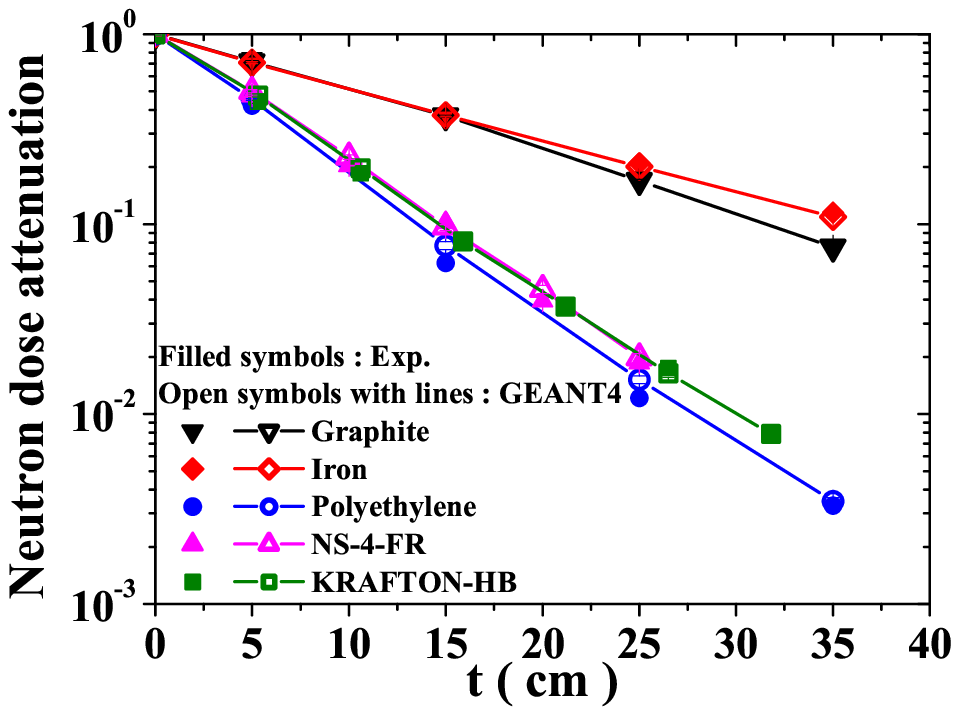, width=6in}
\caption{(Color online) Neutron dose attenuations 
for graphite, iron, polyethylene, NS-4-FR and KRAFTON-HB 
are plotted as a function of thickness t. 
The filled symbols and open symbols with lines denote the experimental 
and the calculated neutron dose attenuations, respectively.}
\label{figure10}
\end{figure}

Figure \ref{figure10} shows the neutron dose attenuations 
for five shielding materials considered in this work.
The filled symbols and open symbols with lines denote the experimental 
and the calculated neutron dose attenuations, respectively.
Only the results from GEANT4 are denoted in this figure, 
because differences between the results from GEANT4 and PHITS 
are not significant.
Both the experiments and our simulations show the same trend in this figure.
As mentioned before, 
the calculated and the experimental values for both graphite and iron 
almost overlap with each other within $\sim$5\% errors.
The differences between the calculated and the experimental values for 
polyethylene, NS-4-FR and KRAFTON-HB are within $\sim$30\%, $\sim$10\% 
and $\sim$10\% errors, respectively. 

%%%%%%%%%%%%%%%%%%%%%%%%%%%%%%%%%%%%%%%%%%%%%%%%%%%%%%%%%%%%%%%%%%%%%%%%%%%%%
%%%%%%%%%%%%%%%%%%%%%%%%%%%%%%%%%%%%%%%%%%%%%%%%%%%%%%%%%%%%%%%%%%%%%%%%%%%%%

In Fig. \ref{figure10}, 
it can be seen that both graphite and iron materials reduce the neutron dose 
up to one order of magnitude when the thickness is 35 cm.
Even though the energy distributions of the neutrons for graphite 
differ from those for iron as can be seen in Fig. \ref{figure4},
the neutron dose attenuation for graphite is almost the same as 
that for iron at t = 5 cm and 15 cm.
As the thickness of the shield increases, however, 
difference between the attenuations for graphite and iron become significant.
When polyethylene, NS-4-FR and KRAFTON-HB shielding materials are used, 
doses are reduced very effectively compared to graphite and iron.
Shielding abilities of polyethylene, NS-4-FR and KRAFTON-HB with 15 cm of thickness 
are comparable to or better than those of graphite and iron with 35 cm of thickness.
When the thickness of the shielding materials is larger than 25 cm, 
neutron doses are reduced by two orders of magnitudes.

\section{Summary}

Neutron dose equivalent rates for various shielding materials are calculated 
by using GEANT4 and PHITS code.
As a neutron source, $^{252}$Cf is assumed.
Five different shielding materials such as graphite, iron, polyethylene, 
NS-4-FR and KRAFTON-HB are considered.
For low energy neutron interactions, 
G4HP models with G4NDL 4.2 based on ENDF/B-VII 
and JENDL-4.0 library are used for GEANT4 
and PHITS calculations, respectively.

First, the neutron energy distributions scored 
in the detector region with and without shielding materials are calculated.
The results obtained from GEANT4 and PHITS are compared with each other.
Also, an old version of GEANT4 and G4NDL 3.13 based on ENDF/B-VI 
as in \cite{bench8} are considered and the results are compared.
It is found that differences between the calculations 
from GEANT4 and PHITS and the difference between the old and new versions of GEANT4 
are not significant for all the cases considered in this work, 
which shows reliability of these Monte Carlo simulations.

Second, the neutron dose equivalent rates with the calculated neutron spectra 
and conversion factors are obtained.
The results are compared with the experimental dose equivalent rates \cite{ueki3, ueki4} 
and the values calculated earlier \cite{bench1, bench3, bench5}.
From the comparison, it is found that GEANT4 and PHITS results 
based on nuclear data libraries describe the 
experimental dose equivalent rates quite well for the graphite, 
iron, NS-4-FR and GRAFTON-HB with the discrepancy 
between the experimental dose equivalent rates 
and the calculated values less than $\sim$20\%. 
However, in the case of polyethylene, 
the discrepancy is up to $\sim$40\%.
Other studies \cite{bench1, bench3, bench6}
as well as the present study show consistent features.

Neutron dose attenuations for the shields are also obtained.
For graphite or iron shielding materials, 
the simulation results are consistent with the experiments 
within $\sim$5\% errors.
The differences between the simulations and the experiments for 
polyethylene, NS-4-FR and KRAFTON-HB are less than 
$\sim$30\%, $\sim$10\% and $\sim$10\%, respectively.

\begin{acknowledgments}
This work was supported in part by the Basic
Science Research Program through the Korea Research
Foundation (NRF-2011-0025116, NRF-2012R1A1A2007826, NRF-2012M2B2A4030183).
\end{acknowledgments}

%\begin{references}

%\end{references}

%\section{Table}

%\section{Figures}

\end{document}